\documentclass[aps,pre,twocolumn,preprintnumbers,showpacs,superscriptaddress]{revtex4}

\begin{document}

\title{Fisher Waves and Front Roughening in a Two-Species Invasion
Model with Preemptive Competition}

\author{L. O'Malley}
\email{omalll@rpi.edu}
\affiliation{Department of Physics, Applied
Physics, and Astronomy, Rensselaer Polytechnic Institute, 110 8th
Street, Troy, NY 12180-3590, USA}

\author{B. Kozma\footnote{Permanent address: Laboratoire de Physique Th\'eorique
(UMR du CNRS 8627), B\^atiment 210 Universit\'e Paris-Sud - 91405
Orsay Cedex, France}}
\email{kozmab@th.u-psud.fr}
\affiliation{Department of Physics, Applied Physics, and
Astronomy, Rensselaer Polytechnic Institute, 110 8th Street, Troy,
NY 12180-3590, USA}

\author{G. Korniss}
\email{korniss@rpi.edu}
\affiliation{Department of Physics, Applied Physics, and Astronomy,
Rensselaer Polytechnic Institute, 110 8th Street, Troy, NY 12180-3590, USA}

\author{Z. R\'acz}
\email{racz@general.elte.hu}
\affiliation{Institute for Theoretical Physics - HAS, E\"otv\"os University,
P\'azm\'any s\'et\'any 1/a, 1117 Budapest, Hungary}

\author{T. Caraco}
\email{caraco@albany.edu}
\affiliation{Department of Biological Sciences, University at Albany,
Albany NY 12222, USA}

\begin{abstract}
We study front propagation when an invading species competes with
a resident; we assume nearest-neighbor preemptive competition for
resources in an individual-based, two-dimensional lattice model.
The asymptotic front velocity exhibits power-law dependence on the
difference between the two species' clonal propagation rates (key
ecological parameters).  The mean-field approximation behaves
similarly, but the power law's exponent slightly differs from the
individual-based model's result.  We also study roughening of the
front, using the framework of non-equilibrium interface growth. Our
analysis indicates that initially flat, linear invading fronts
exhibit Kardar-Parisi-Zhang (KPZ) roughening in one transverse
dimension. Further, this finding implies, and is also confirmed by
simulations, that the temporal correction to the asymptotic front
velocity is of ${\cal O}(t^{-2/3})$.
\end{abstract}


\date{\today}
\maketitle

\section{Introduction}

The dynamics of propagating fronts integrates concepts shaping our
understanding of how invasive species, emerging infectious
disease, and evolutionary adaptations spread across ecological
landscapes \cite{MURRAY_03}. Indeed, objects as seemingly
dissimilar as chemical reaction fronts \cite{Riordan_PRL95} and
the spread of opinions \cite{EBN_05} share certain basic
spatio-temporal properties.  Fisher \cite{FISHER_37} and
Kolmogorov et al. \cite{KOLMOGOROV_37} first addressed velocity
characteristics of simple fronts by modeling a favored mutation's
one-dimensional spread with a reaction-diffusion equation.  A
lengthy series of biologically generalized reaction-diffusion
models has since appeared \cite{Metz}. However, recent
developments emphasize the ecological realism of discrete
individuals \cite{Durett94b,KC_JTB,OBYKAC_05}. Our study
analyzes front propagation when two plant species compete
preemptively
\cite{SHURIN_04,AMARA_rev_03,TANEY_00,YU_01,Oborny_05} for a
common limiting resource. Discrete individuals of each species
propagate clonally, so that competitive interactions are spatially
localized.  An ``invader" species has an individual-level
reproductive advantage over a ``resident" species, so that
competition is asymmetric.

This paper focuses on one-dimensional fronts separating the
species in a two-dimensional environment.  We study the asymptotic
front velocity, as well as the temporal and finite-size
corrections (or rates of convergence) to this velocity.
Furthermore, we investigate roughening of the moving fronts from a
non-equilibrium interface viewpoint
\cite{Barabasi_FCSG,Zhang_review}. Asymptotic properties of
initially flat, linear fronts do offer insights concerning the
competitive dynamics of locally propagating plants. Consider
trembling aspen ({\it Populus tremuloides}).  Significant seed
production, hence long-distance dispersal, occurs only about once
each five years \cite{Sakai}.  Most growth is clonal, where a new
tree grows from an existing tree's roots.  Clonally propagated
trembling aspen clusters occasionally expand to several thousand
trees, and cover $>40$ ha \cite{Kemperman}.  To model such
systems, one can assume that introductions of an invader by seed
occur rarely and stochastically, in both space and time. We have
shown \cite{KC_JTB,OBYKAC_05,YKC_05,OAKC_SPIE} that the time
evolution of the invader and resident populations in such systems
can be well described within the framework of homogeneous
nucleation and growth \cite{avrami}. In particular, in two
dimensions, for sufficiently large systems, the typical time
(lifetime) until competitive exclusion of the weaker competitor
scales as $\tau$$\sim$$(Iv^2)^{-1/3}$
\cite{KC_JTB,OBYKAC_05,YKC_05,OAKC_SPIE}, where $I$ is the
stochastic nucleation rate per unit area of the successful
clusters of the better competitor, and $v$ is the asymptotic
radial velocity of the growing (on average) circularly symmetric
fronts. It is, thus, clear that the full understanding of the
dependence of the lifetime on the local rates of the systems
requires the knowledge of the velocity of the front separating the
two species. Furthermore, as circular fronts grow sufficiently
large, so that curvature corrections become negligible, frontal
velocities approach values for linear fronts \cite{MURRAY_03}.

Recently we have reported preliminary results on the front
velocity in the model studied here \cite{OKKRT_UGA06}. This paper
extends our analysis by investigating the front's propagation as a
non-equilibrium interface \cite{Barabasi_FCSG,Zhang_review}. Our
Monte Carlo simulations not only provide numerical estimates for
the front velocity, but, through a detailed finite-size analysis,
also identify the universality class of the moving and roughening
interface \cite{Barabasi_FCSG,Zhang_review} separating the two
species. Our results indicate that the asymptotic front velocity
scales as a power law with difference between the two species'
clonal propagation rates. Further, we find that initially flat,
linear invading fronts exhibit Kardar-Parisi-Zhang (KPZ)
\cite{KPZ_86} roughening in one transverse dimension. This finding
implies, and was also confirmed by our simulations, that the
temporal correction to the asymptotic front velocity is of ${\cal
O}(t^{-2/3})$.

We organize the remainder of the paper as follows. In Sec.~II we
define the spatially explicit, individual-based model of
two-species competition. In Sec.~III we compare simulation results
for the asymptotic front velocity with results from the mean-field
equations. In Sec.~IV we carry out the analysis of the interface
roughening characteristics of the front, also yielding the
temporal and finite-size corrections to the asymptotic front
velocity. We discuss and summarize our results in Sec.~V.

\section{Two-Species Invasion Model with Preemptive Competition}

Our analysis treats the velocity and roughening of invading fronts
as functions of the propagation and mortality rates of each
species, with possible temporal and finite-size corrections.  On a
two-dimensional $L_x$$\times$$L_y$ lattice, a site represents the
minimal level of resources necessary to maintain a single
individual. Competition for the resource is preemptive
\cite{SHURIN_04}; that is, a currently occupied site cannot be
colonized by any species until mortality of the occupant opens the
site. The local occupation number at site ${\bf x}$ is $n_i({\bf
x})=0,1$ with $i= 1,2$, representing the number of resident and
invader individuals, respectively. Since two species cannot
simultaneously occupy the same site, the excluded volume
constraint yields $n_1({\bf x})n_2({\bf x}) = 0$. A species may
occupy new sites only through local clonal propagation. Therefore,
a species occupying site ${\bf x}$ may only reproduce if one or
more of its neighboring sites is empty (here we consider only
nearest-neighbor interactions).

During our time unit, one Monte Carlo step per site [MCSS], $L_xL_y$ sites
are chosen at random for updating. The local configuration of a
chosen site is updated according to the following transition
rates. An empty site may be occupied by species $i$ through clonal
reproduction from a neighboring site at rate $\alpha_i\eta_i({\bf
x})$, with $\alpha_i$ being the individual-level reproduction rate
for species $i$, and $\eta_i({\bf x}) = (1/4)\Sigma_{{\bf
x}'\epsilon {\rm nn}({\bf x})}n_i({\bf x}')$ is the density of
species $i$ in the neighborhood around site ${\bf x}$; ${\rm
nn}({\bf x})$ is the set of nearest neighbors of site ${\bf x}$.
An occupied site opens through mortality of the individual; the
mortality rate $\mu$ is the same for both species. The transition
rules for an arbitrary site ${\bf x}$ can be summarized as
follows:
\begin{equation}
0\stackrel{\alpha_1\eta_1(\bf{x})}{\longrightarrow}1, \;\;
0\stackrel{\alpha_2\eta_2(\bf{x})}{\longrightarrow}2, \;\;
1\stackrel{\mu}{\longrightarrow}0, \;\;
2\stackrel{\mu}{\longrightarrow}0, \;\;
\label{rates}
\end{equation}
where 0, 1, 2 indicates whether a site is open, resident-occupied,
or invader occupied, respectively.

One should note that the above discrete stochastic model, governed
by preemptive competition, is a two-species generalization
\cite{OBYKAC_05} of the basic contact process
\cite{Durett94b,Harris_74,Oborny_05}. Each species, in the absence
of the other one, becomes extinct (through a transition to an
absorbing phase {\cite{Hinrichsen_00}) if
$\alpha_i$$<$$\alpha_c(\mu)$, where $\alpha_c(\mu)\approx 1.65\mu$
\cite{Oborny_05,OBYKAC_05} [and $\alpha_c(\mu)=\mu$ in the
mean-field approximation, see Sec. III.A]. We are interested in
the scenario where {\em competition} between the two species
drives the dynamics (i.e., not extinction by insufficient
colonization rates), and one of the species (the invader) has a
reproductive advantage over the resident. Hence, we study the
$\alpha_c(\mu)<\alpha_1<\alpha_2$ regime.

To study front propagation, we impose periodic boundary conditions
along the $y$-direction of an $L_x$$\times$$L_y$ lattice. The
initial condition is a flat linear front (straight vertical line),
i.e., the invader completely occupies a few vertical columns at
the left edge of the lattice, and all other remaining sites are
occupied by the resident species. The direction of propagation,
therefore, is along the $x$-direction. As the simulation begins,
many individuals of both species die in a few time steps, and the
densities on both sides of the front quickly relax to their
``quasi-equilibrium'' values, where clonal propagation is balanced
by mortality [Fig.~\ref{fig1}]. As the simulation evolves, we
track the location of the invading front by defining the edge as
the location of the right-most individual of the invading species,
$h_y(t)$, for each row $y$. The average position
$\overline{h}(t)=(1/L_y)\sum_{y}h_y(t)$ is then recorded for each
time step, from which we extract the velocity [as
$\overline{h}(t)$ approaches a constant slope for late times]. We
ran each simulation until the front reached the end of the system.
The longitudinal system size $L_x$ has no particular impact on the
system's time evolution, although it constrained the maximal
length of our simulations.

One can also observe [Fig.~\ref{fig1}], that as the front
propagates, it also ``roughens'', i.e., the typical size of the
fluctuations about the mean front position is increasing, before it
reaches the steady-state for a given transverse system size $L_y$. This
kinetic roughening phenomenon \cite{Barabasi_FCSG,Zhang_review} will be discussed
in detail in Sec. IV.
\begin{figure}[t]
  \centering
  \vspace*{3.00truecm}
  \includegraphics{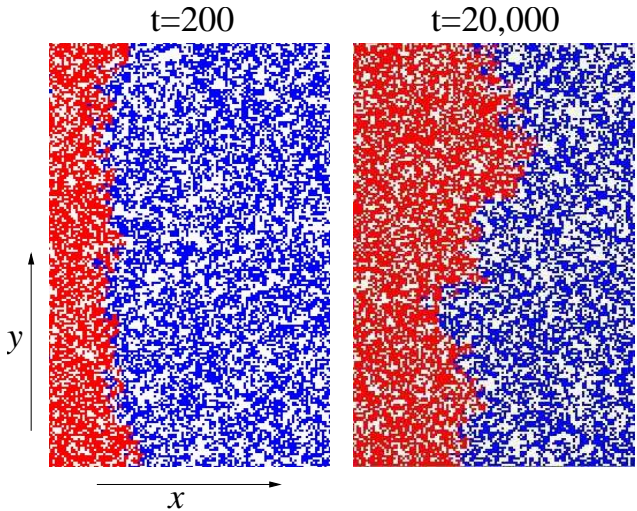}
  \vspace*{4.00truecm}
\caption{Snapshots of the moving and roughening front in the
early-time regime ($t$$=$$200$ MCSS) and in the steady state
($t$$=$$20,000$ MCSS) for $\alpha_1$$=$$0.50$,
$\alpha_2$$=$$0.70$, $\mu$$=$$0.20$, and $L_y$$=$$200$. White
represents empty sites, while blue and red correspond to sites
occupied by the residents and the invaders, respectively.}
\label{fig1}
\end{figure}

\section{Invasion as Propagation into an Unstable State}

\subsection{Mean-field equations and the asymptotic front velocity}

Taking the master equation corresponding to transition rates in
Eq.~(\ref{rates}), and neglecting correlations between densities
at different sites, yields dynamics of the ensemble-averaged local
densities $\rho_i({\bf x},t)$$\equiv$$\langle n_i({\bf
x},t)\rangle$. We obtain
\begin{eqnarray}
\rho_i({\bf x},t+1) - \rho_i({\bf x},t) & = & \left[1 -
\rho_1({\bf x},t) - \rho_2({\bf x},t)\right]
\nonumber \\
\times \frac{\alpha_i}{4}\sum_{{\bf x'}\epsilon {\rm nn}({\bf
x})}\rho_{i}({\bf x'},t)
 & - & \mu \rho_i({\bf x}, t) \;.
\label{mfe}
\end{eqnarray}
where $i = 1,2$. Taking the naive continuum limit of the above
equations, one obtains the (coarse-grained) equations of motion
\begin{eqnarray}
\partial_t \rho_i({\bf x},t) & = &
\frac{\alpha_i}{4} \left[1 - \rho_1({\bf x},t) - \rho_2({\bf
x},t)\right] \nabla^2\rho_i({\bf x},t)
\nonumber\\
& + &  \alpha_i \left[1 - \rho_1({\bf x},t) - \rho_2({\bf
x},t)\right] \rho_i({\bf x},t)\nonumber \\ & - & \mu \rho_i({\bf x},t)\;,
\label{cmfe}
\end{eqnarray}
$i=1,2$. The spatially homogeneous solutions of these equations,
$(\rho_1^*,\rho_2^*)$, are $(0,0), (1 - \mu/\alpha_1,0)$, and
$(0,1 - \mu/\alpha_2)$. In the parameter regime of our interest,
$\mu < \alpha_1 < \alpha_2$, only the last solution $(0,1 -
\mu/\alpha_2)$ is stable.  Thus, the propagation of a front
separating the stable $(0,1 -\mu/\alpha_2)$ (invader dominated)
and unstable $(1 - \mu/\alpha_1,0)$ (resident dominated) regions
amounts to propagation into an unstable state
\cite{FISHER_37,KOLMOGOROV_37, Aronson_78, Dee_83, Saarloos_87}, a
phenomenon that has generated a vast amount of literature
\cite{Saarloos_03} since the original papers by Fisher
\cite{FISHER_37} and Kolmogorov et al. \cite{KOLMOGOROV_37}. At
the level of the mean-field equations, the front is ``pulled'' by
the leading edge into the unstable state. Then, for a sufficiently
sharp initial density profile \cite{MURRAY_03,Saarloos_03}, the
asymptotic velocity is determined by the infinitesimally small
density of invaders that intrude into the linearly unstable region
dominated by the resident species. Linearizing Eqs.~(\ref{cmfe})
about the unstable state, $\rho_1=1-\mu/\alpha_1 + \phi_1$,
$\rho_2=\phi_2$, one immediately obtains for the density of
invaders
\begin{equation}
\partial_t \phi_2({\bf x},t) =
\frac{\mu}{4}\frac{\alpha_2}{\alpha_1}\nabla^2\phi_2({\bf x},t)
+ \mu\left(\frac{\alpha_2}{\alpha_1} - 1\right)\phi_2({\bf x},t) \;.
\label{lcmfe}
\end{equation}
Performing standard analysis
\cite{MURRAY_03,Dee_83,Saarloos_87,Saarloos_03} on the above equation,
we obtain the asymptotic velocity of the ``marginally stable'' invading fronts
\begin{equation}
v^{*} = \frac{\mu}{\alpha_1} \sqrt{\alpha_2 (\alpha_2 - \alpha_1)}.
\label{velocity}
\end{equation}
The velocity $v^*$ above is the minimum velocity of a physically
allowed travelling wave, permitted by Eqs.~(\ref{lcmfe}), and is
actually realized by {\em deterministic} nonlinear
reaction-diffusion dynamics for sufficiently sharp initial
profiles \cite{MURRAY_03,Dee_83,Saarloos_87,Saarloos_03}. For
further comparisons, we also note that the above asymptotic front
velocity is approached as $v(t)=v^{*}-{\cal O}(1/t)$
\cite{Saarloos_03}. Also, as can be seen from
Eq.~(\ref{velocity}), for small differences in the reproduction
rates of the two species (a parameter of ecological significance),
the front velocity scales as
$v^{*}$$\sim$$(\alpha_2-\alpha_1)^{\theta}$ with $\theta=1/2$.
\begin{figure}[t]
\centering
\vspace*{2.50truecm}
       \includegraphics{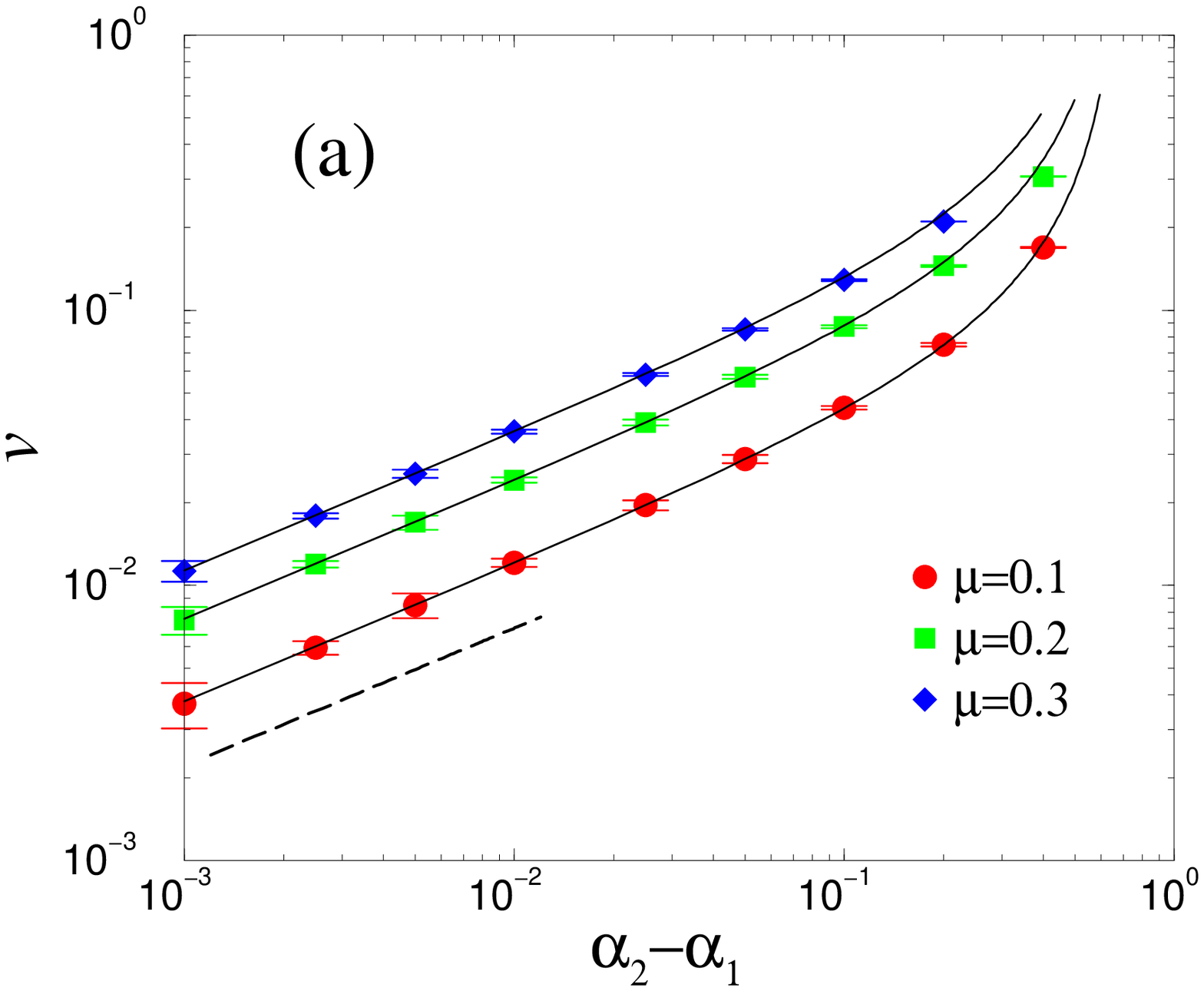}
\vspace*{5.80truecm}
       \includegraphics{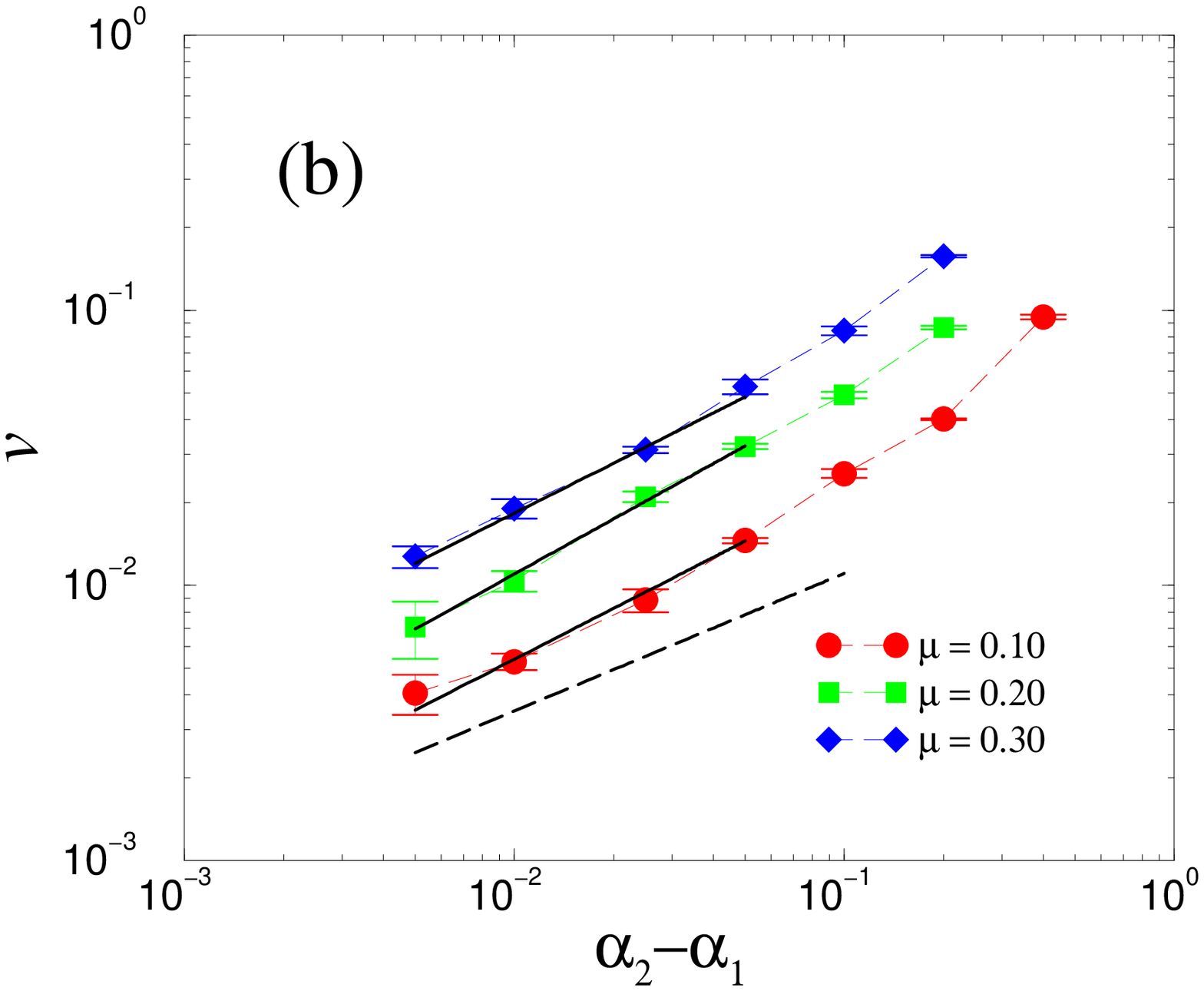}
\vspace*{3.5truecm}
\caption{(Color online)
(a) Asymptotic mean-field front velocities for three different mortality
rates for fixed $\alpha_2$=$0.70$, as a function of the difference of propagation
rates, $\alpha_2$$-$$\alpha_1$. Symbols are obtained by
numerically iterating the nonlinear mean-field equations of motion [Eq.~(\ref{mfe})].
The solid curves running through the data points are the analytically
obtained velocities in the ``leading edge'' approximation
[Eq.~(\ref{velocity})], as described in the text. The straight dashed
line indicates the slope $\theta$$=$$0.5$, corresponding to the exponent
of an effective power law for small differences between the reproduction rates.
(b) Asymptotic front velocities from Monte Carlo simulations for the individual-based
stochastic model for fixed
$\alpha_2$=$0.70$ as in (a), with $L_{y}=100$, for three
values of $\mu$. The solid straight lines are the best-fit effective
power laws in the region where the difference between the reproduction
rates is small, corresponding to $\theta=0.61\pm0.04$,
$\theta=0.66\pm0.04$, and $\theta=0.60\pm0.04$ for $\mu$$=$$0.1$,
$\mu$$=$$0.2$, and $\mu$$=$$0.3$, respectively.
For comparison, the straight dashed line corresponds to the effective power law of the
mean-field case with an exponent $\theta$=$0.5$.}
\label{fig2}
\end{figure}

It is important to note that the front velocity given by
Eq.~(\ref{velocity}), obtained by linearizing Eq.~(\ref{cmfe}),
{\em fully} reproduces the velocity obtained by numerically
iterating the {\em non-linear}  continuous-density mean-field
equations [Eq.~(\ref{mfe})], as can be seen in Fig.~\ref{fig2}(a).
This is a generic and powerful feature of deterministic pulled
fronts where, despite the non-linearity of the full dynamics,
velocities are completely determined by the leading edge
\cite{MURRAY_03,Saarloos_03}.

\subsection{Monte Carlo results for the front velocity}
We now present results for the discrete individual-based
stochastic model defined by the transition rates
Eq.~(\ref{rates}). We found that the front velocity is much
smaller than that of the mean-field approximation as shown in
Fig.~\ref{fig2}(b) [cf. Fig.~\ref{fig2}(a)]. Further, for small
differences in the reproduction rates, the velocity is described
by an effective power law
$v^{*}$$\sim$$(\alpha_2-\alpha_1)^{\theta}$ with $\theta
\stackrel{\sim}{>} 0.6$ [Fig.~\ref{fig2}(b)], slightly, but
distinctly, greater than the exponent $\theta$$=$$0.5$ of the mean-field
case. Similar deviations from the mean-field exponent have been
found in two-dimensional epidemic \cite{Warren_01} and
reaction-diffusion models \cite{Moro_01}. The discreteness of the
individuals \cite{Brunet_97,Kessler_98a,Kessler_98b} and noise
\cite{Doering_03,Doering_05} have been shown to contribute to a
velocity that is different from the mean-field description. The
general belief is that fronts in stochastic individual-based (or
particle) models are ``pushed'', in the sense that the front
velocity is determined by the full non-linear front region,
instead of an infinitesimally small leading edge
\cite{Saarloos_03}, a behavior predicted by the mean-field
approximation. Our two-species model provides an example for this
generic behavior.

An interesting feature of the invasion fronts is that their
propagation velocity approaches the asymptotic value rather slowly
(e.g. as ${\cal O}(1/t)$ in mean-field models \cite{Saarloos_03}).
Thus, from application point of view, it is important to establish
how the front reaches its asymptotic velocity. In the next
section, we are going to analyze both the temporal and finite-size
approach toward the asymptotic front velocity, along with other
universal characteristics (such interface roughening
\cite{Barabasi_FCSG,Zhang_review}) of the model, using the
framework of scaling in non-equilibrium interfaces.

\section{Interface Roughening and Corrections to the Asymptotic Front Velocity}

\subsection{Dynamic scaling}

To extract the scaling properties of the roughening of the front, as
can be seen in Fig.~\ref{fig1}, we analyze the width of the front
\begin{equation}
\langle w^2(L_y,t)\rangle = \left\langle\frac{1}{L_y}\sum_{y = 1}^{L_y}
[h_y(t) -\overline{h}(t)]^2\right\rangle \;,
\label{w2}
\end{equation}
where $h_y(t)$ is defined as the location of the leading individual
in row $y$ for the invading species. In what follows, we define
$L\equiv L_y$, and investigate the scaling properties of the width
$\langle w^2(L,t)\rangle$ within the standard framework of dynamic
scaling of non-equilibrium interfaces \cite{Barabasi_FCSG,Zhang_review}.

The temporal and system-size scaling of the width $\langle
w^2(L,t)\rangle$ typically identifies the universality class of
the growing front. In finite systems, the width grows as $\langle
w^2(L,t)\rangle$$\sim$$t^{2\beta}$ from early to intermediate
times. At a system-size-dependent crossover time,
$t_\times$$\sim$$L^z$, it saturates (reaches steady state) and
scales as $\langle w^2_{sat}\rangle \equiv\langle
w^2(L,\infty)\rangle\sim L^{2\alpha}$, where $L$ is the transverse
linear system size. $\alpha$, $\beta$, and  $z$ are referred to as
the roughness, growth, and dynamic exponents, respectively.
Further, these exponents are not independent, but obey the scaling
relation $\alpha=\beta z$. The above temporal and system-size
behavior, with the appropriate crossover time, can be captured by
the Family-Vicsek scaling form \cite{FV}
\begin{equation}
\langle w^2(L,t)\rangle = L^{2\alpha}f(t/L^z).
\label{FV}
\end{equation}
For small values of its argument, $f(x)$ behaves as a power law, while
for large arguments it approaches a constant
\begin{equation}
f(x)=\left\{
\begin{array}{ll}
x^{2\beta}       & \mbox{for $x \ll 1$} \\
{\rm const.}     & \mbox{for $x \gg 1$}
\end{array} \right. \;,
\label{f_x}
\end{equation}
yielding the scaling behavior of the width in the growth and
steady-state regime, provided the scaling relation for the exponents,
$\alpha=\beta z$, holds.
\begin{figure}[t]
\centering
\vspace*{2.50truecm}
       \includegraphics{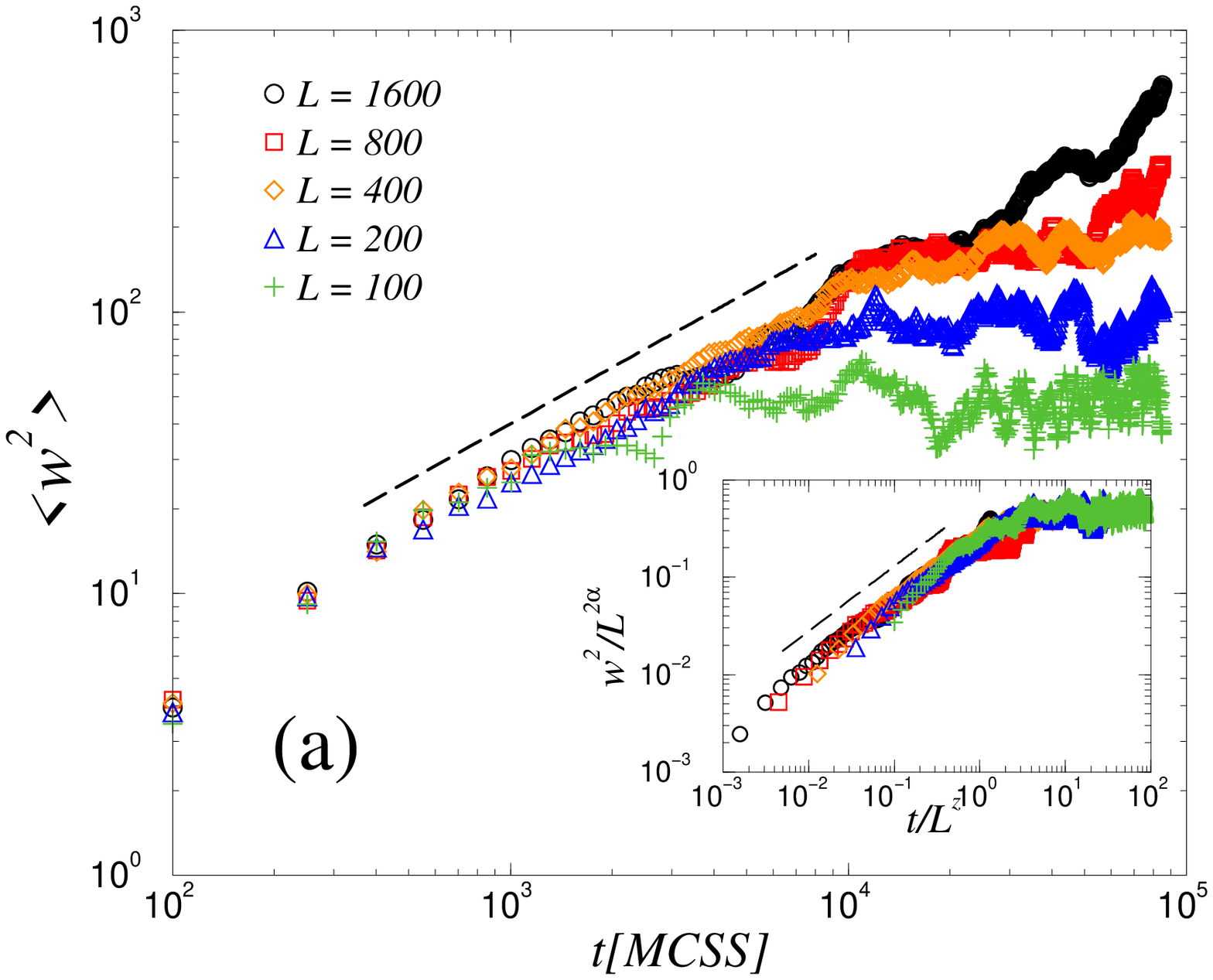}
\vspace*{5.80truecm}
       \includegraphics{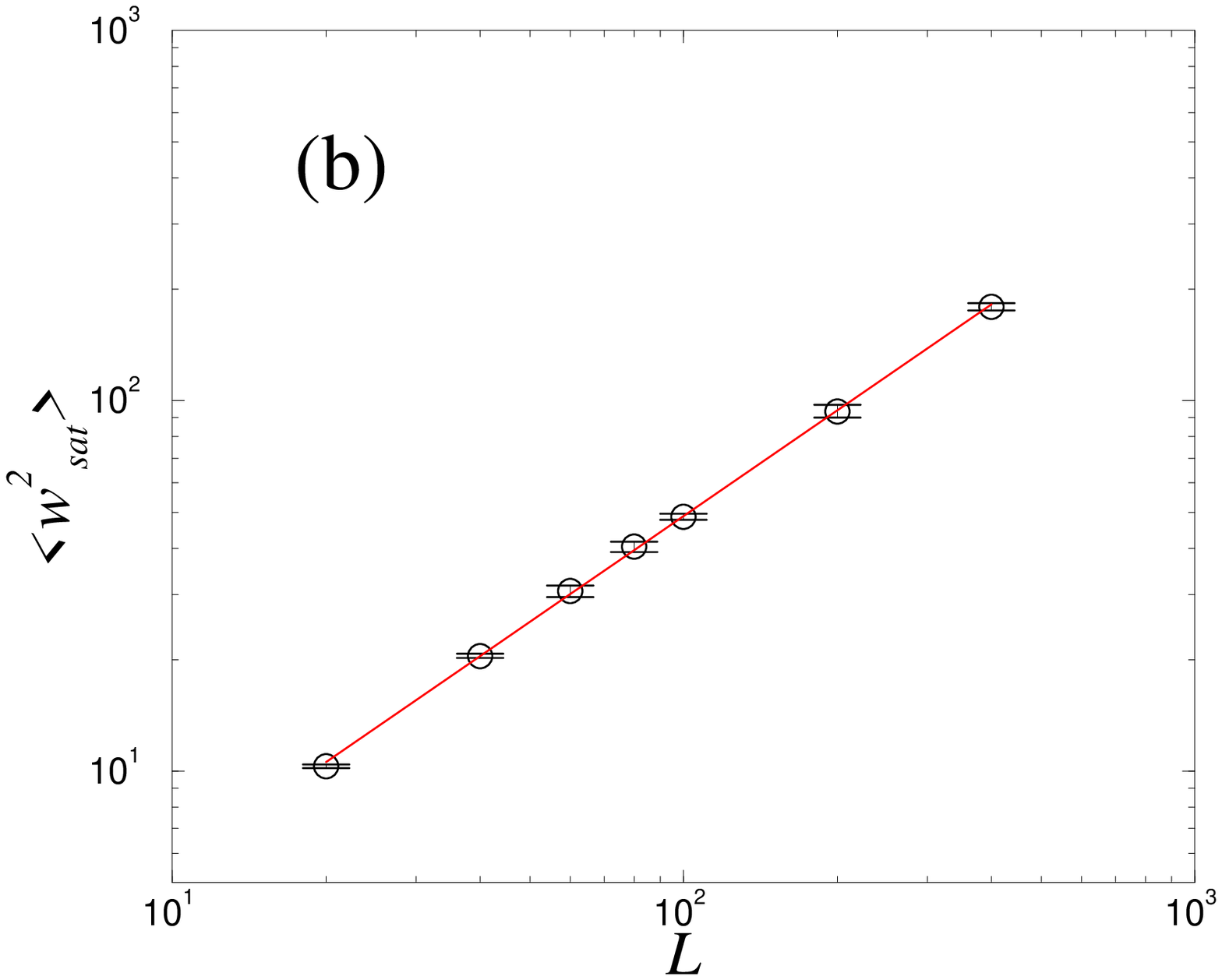}
\vspace*{3.0truecm}
\caption{(Color online)
       (a) The average width as a function of time
       (on log-log scales) for various
       system sizes, averaged over $20$ independent realizations, for
       $\alpha_1=0.70$, $\alpha_2=0.80$, and $\mu=0.10$.
       The dashed line corresponds to the one-dimensional KPZ power law
       with the exponent $2\beta=2/3$. The inset shows the scaled plot,
       $\langle w^2(L,t)\rangle/L^{2\alpha}$ vs. $t/L^z$
       using the one-dimensional KPZ exponents.
       (b) Steady-state width as a function of the system size,
       $L (\equiv L_y)$, for the same clonal propagation and mortality rates as in (a).
       The solid line corresponds to the best-fit power law with the
       exponent $2\alpha =0.95\pm0.01$.}
\label{fig3}
\end{figure}

As as shown in Figs.~\ref{fig3}, our results show reasonable
agreement with the exponents of the well-known KPZ universality
class ($\beta$$=$$1/3$, and $\alpha$$=$$1/2$)
\cite{Barabasi_FCSG,Zhang_review,KPZ_86}. Further, the scaled
width, $\langle w^2(L,t)\rangle/L^{2\alpha}$ vs. $t/L^z$ produces
good data collapse, as suggested by Eq.~(\ref{FV}), and confirms
dynamic scaling for the invasion process [Fig.~\ref{fig3}(a)
inset].

\subsection{Steady-state width distribution}

For further analysis, we also constructed the full distributions
$P(w^2)$ of the steady-state width for different system sizes [i.e.,
the normalized histograms of the width
obtained from the steady-state time series]. This observable typically provides an
additional strong signature of the underlying universality
class of the fluctuating and growing interface \cite{foltin}. In
particular, for the one-dimensional KPZ class, it has been obtained
analytically \cite{foltin} and can be written in the generic scaling form
$P(w^2) = \langle w^2 \rangle^{-1} \Phi(w^2/\langle w^2 \rangle)$ with
\begin{equation}
\Phi(x) = \frac{\pi^2}{3}\displaystyle\sum_{n = 1}^\infty (-1)^{n-1}
n^2 e^{-(\pi^2 / 6) n^2 x}.
\label{P_w}
\end{equation}
\begin{figure}[t]
\centering
\vspace*{2.50truecm}
       \includegraphics{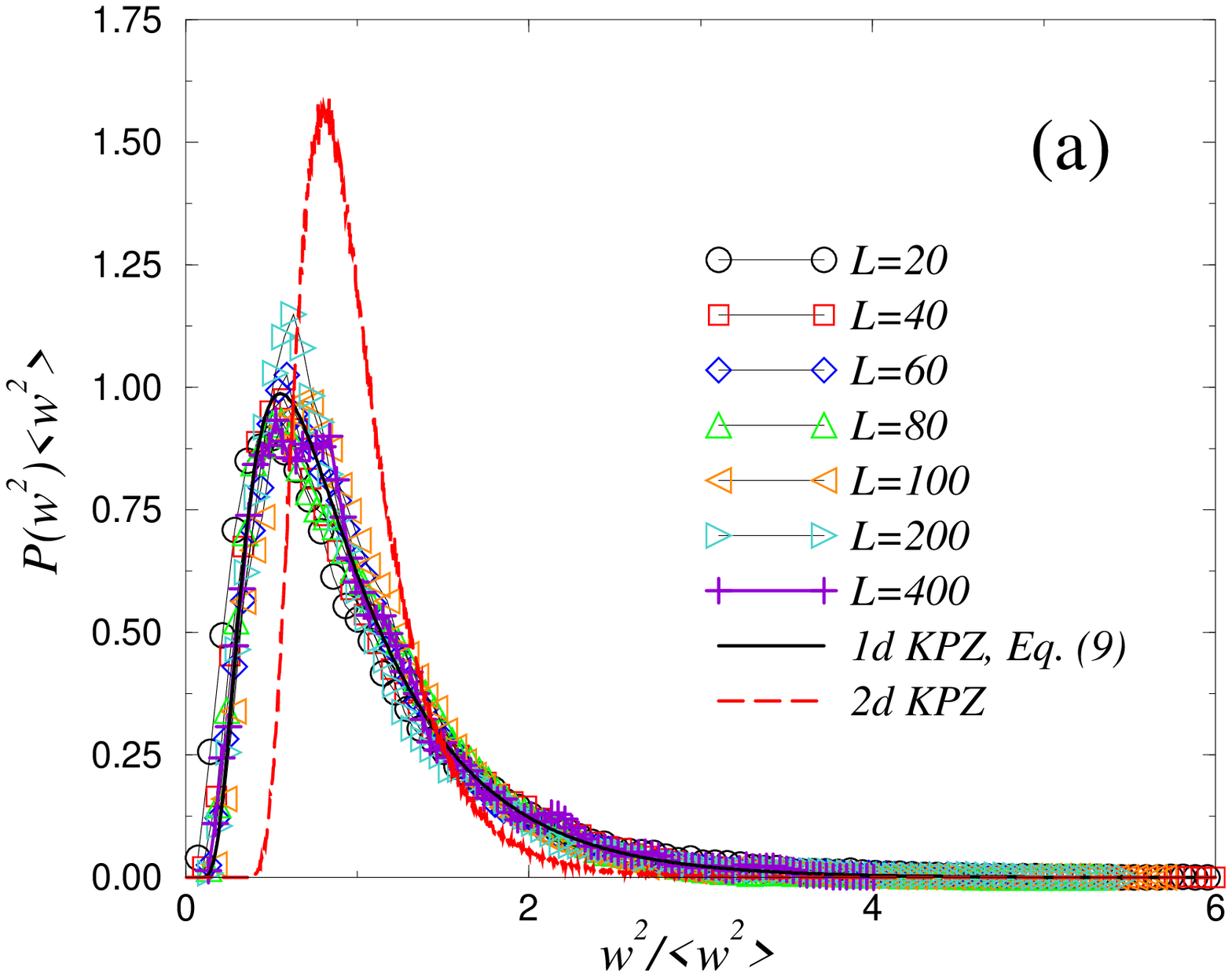}
\vspace*{5.80truecm}
       \includegraphics{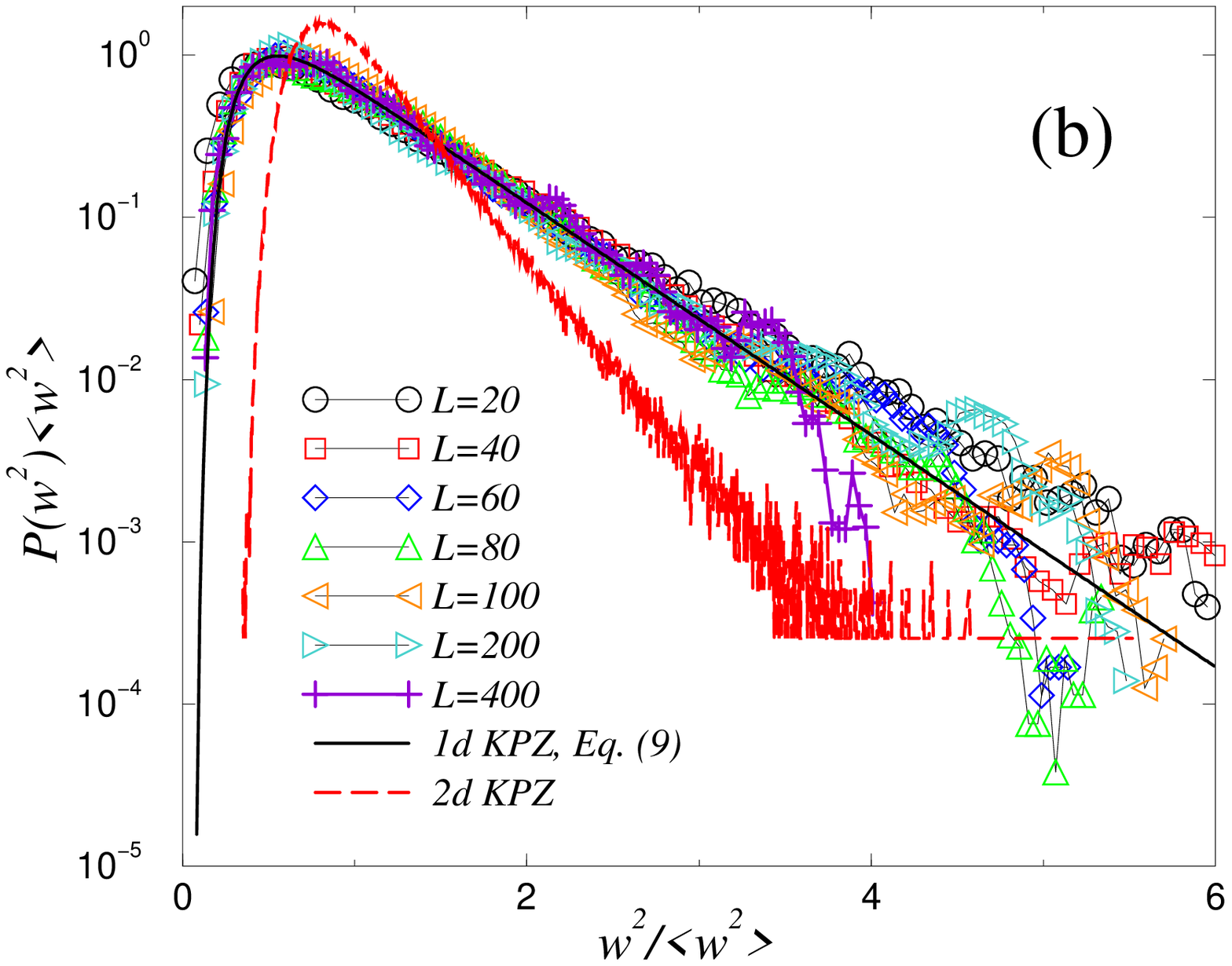}
\vspace*{3.5truecm}
\caption{(Color online)
(a) Steady-state width distributions for $\alpha_1$$=$$0.70$,
$\alpha_2$$=$$0.80$, and $\mu$$=$$0.10$ for various system sizes $L$. The
solid curve is the scaled analytic width  distribution of the
one-dimensional KPZ class \cite{foltin}, Eq.~(\ref{P_w}). For comparison, the
two-dimensional scaled KPZ width distribution is also shown, as
obtained in Ref.~\cite{Marinari_PRE90}
(b) The same as (a) but on log-normal scales.}
\label{fig4}
\end{figure}

In Fig.~\ref{fig4} we show the scaled width distribution
$P(w^2)\langle w^2 \rangle$ vs. $w^2/\langle w^2 \rangle$ for
various system sizes and compare it with the above analytic
scaling function $\Phi(x)$. Our data, again, strongly suggest that
propagating planar fronts in our two-species invasion model belong
to the one-dimensional KPZ universality class. The large
deviations and data scattered around peaks of the distributions
are due to sampling error.  Steady-state time series for larger
systems become strongly correlated, and so require excessively
long simulations to generate sufficiently large statistically
independent samples. Fig.~\ref{fig4} also shows the KPZ width
distribution  in two transverse dimensions \cite{Marinari_PRE90},
offering a comparison suggested by a recent, somewhat
counterintuitive conjecture \cite{Saarloos_PRL01}. That conjecture
suggests that fronts which are pulled in the mean-field limit,
belong to the one higher dimensional KPZ class than one would
normally expect (i.e., two instead of one transverse dimension in
our model). We give more discussion on this topic aspect in Sec.
V.

\subsection{Temporal and finite-size corrections to the asymptotic
front velocity}

\begin{figure}[t]
\centering
\vspace*{2.50truecm}
       \includegraphics{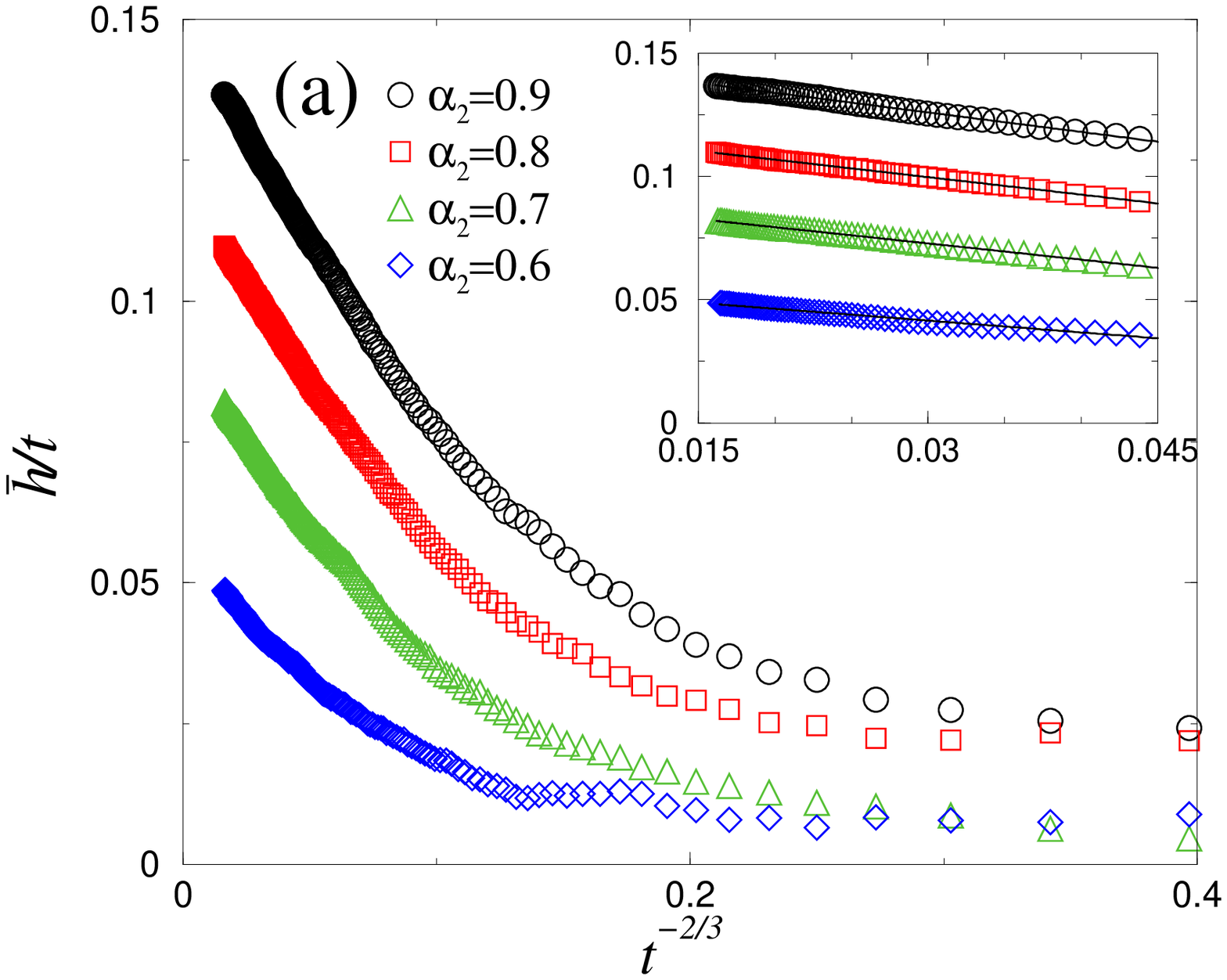}
\vspace*{5.80truecm}
       \includegraphics{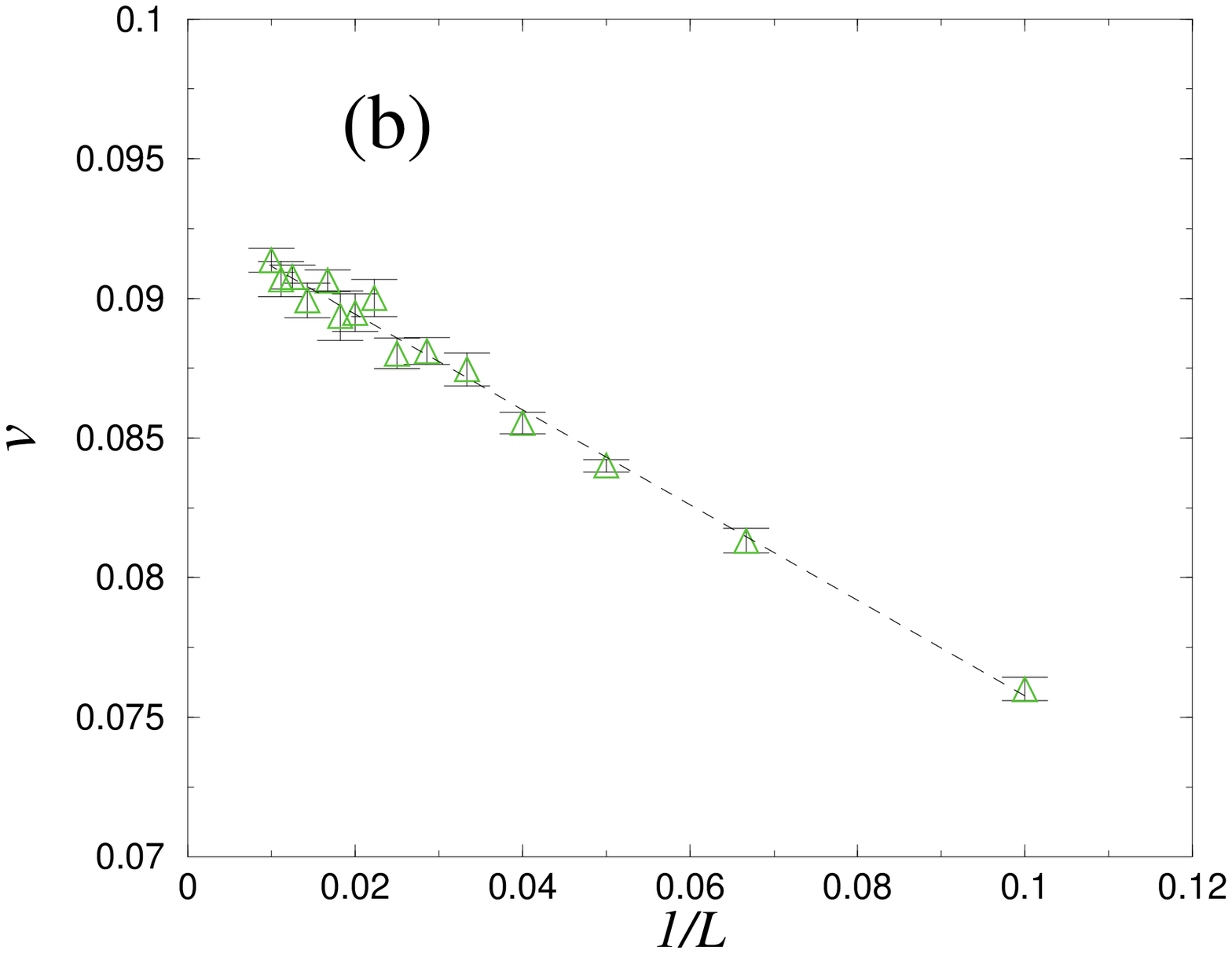}
\vspace*{3.0truecm}
\caption{(Color online)
  (a) Temporal corrections to the asymptotic front velocity
  for $L$$=$$800$ for
  different values of $\alpha_2$, with  $\alpha_1$$=$$0.50$ and $\mu$$=$$0.20$.
  The  horizontal (time) axis is scaled as $t^{-2/3}$, motivated by the form
  of the corrections of the one-dimensional KPZ class [Eq.~(\ref{v_corr})]. The inset
  enlarges the regime of (a) where the universal temporal corrections
  follow the KPZ behavior. The straight solid lines correspond to
  the linear scaling as a function of $t^{-2/3}$.
  (b) Finite-size corrections to the asymptotic velocity as a function of $L^{-1}$,
  motivated by the universal corrections of the one-dimensional KPZ
  class, for $\alpha_1$$=$$0.50$, $\alpha_2$$=$$0.70$, and $\mu$$=$$0.20$.
  The straight dashed line corresponds to the linear behavior as a function of $1/L$.
  }
\label{fig5}
\end{figure}
Although the actual value of the velocity of the invading front is
not universal, Krug and Meakin \cite{Krug_90} showed that the forms of the
temporal and finite-size corrections of the front velocity,
$v(L,t)=d\overline{h}(L,t)/dt$,
are universal. More specifically, corrections to the asymptotic front
velocity $v^*$ are given by  \cite{Krug_90}
\begin{equation}
v(L,t)=
\left \{ \begin{array}{lll}
v^* - c_1 t^{-2(1-\alpha)/z} & \mbox{for} & t \ll L^z \\
v^* - c_2 L^{-2(1-\alpha)}   & \mbox{for} & t \gg L^z
\end{array}\right. \;,
\label{growth_corr}
\end{equation}
where $\alpha$ and $z$ are the roughness and dynamic exponents
characterizing the universality class of the model, and $c_1$ and
$c_2$ are non-universal constants
depending on the parameters and microscopic details of the specific model.
Translating their results to the easily measurable quantity
$\overline{h}(t)/t$ (by integrating the above equation with respect to
time and dividing the result by $t$), one also has
\begin{equation}
\frac{\overline{h}(L,t)}{t} =
\left \{ \begin{array}{lll}
v^* - c_1' t^{-2(1-\alpha)/z} & \mbox{for} & t \ll L^z \\
v^* - c_2 L^{-2(1-\alpha)}    & \mbox{for} & t \gg L^z
\end{array}\right. \;,
\label{v_corr}
\end{equation}
where $c_1'=c_1 z/(z+2\alpha-2)$. In particular, for the
one-dimensional KPZ class, $2(1-\alpha)/z=2/3$ and
$2(1-\alpha)=1$. Thus, the early-time temporal, and late-time
finite-size corrections scale as ${\cal O}(t^{-2/3})$ and ${\cal
O}(L^{-1})$, respectively. Our data for the velocity of the
propagating front in the two species-invasion model follows these
corrections very closely [Fig.~\ref{fig5}], offering additional
evidence that the front belongs to the one-dimensional KPZ class.
Also, note that the temporal relaxation of the front velocity is
in contrast to mean-field results where pulled fronts exhibit
${\cal O}(t^{-1})$ corrections to the asymptotic velocity
\cite{Saarloos_03}.

\section{Discussion and Summary}

We studied front propagation in a two-species model for ecological
invasion with preemptive competition, applicable to clonal plants.
We performed dynamic Monte Carlo simulations using the local
transition rules [Eq.~(\ref{rates})] for initially flat linear
fronts. We found the front velocity significantly smaller than
that of the mean-field approximation. Also, for small differences
between local reproduction rates, the asymptotic front velocity
scales as an effective power law
$v^{*}$$\sim$$(\alpha_2-\alpha_1)^{\theta}$ with $\theta
\stackrel{\sim}{>}0.6$, an exponent slightly, but noticeably
different from the mean-field value $\theta=0.5$.  The
discreteness of the individuals
\cite{Brunet_97,Kessler_98a,Kessler_98b} in our lattice model (or,
the effective density cutoffs in a continuum description) and
noise \cite{Doering_03,Doering_05} have been shown to contribute
to the deviations from the mean-field results. More specifically,
fronts in stochastic individual-based models, which in the
mean-field limit converge to a pulled front solution, are instead
``pushed''. Therefore, the front velocity is determined by the
entire non-linear front region, instead of the infinitesimally
small leading edge alone \cite{Saarloos_03}. Our model is an
example for this type of behavior.

We also investigated the universal features of the roughening of
the propagating front, separating the invaders and the residents.
We found that the front roughening in our two-species invasion
model belongs to the one-dimensional KPZ class. We must also place
our results in the context of a recent conjecture by Tripathy et
al. \cite{Saarloos_PRL01} that propagating fronts, which in the
mean-field limit are ``pulled'', exhibit KPZ scaling on a
$(d_{\perp}$$+$$1)$ dimensional ``substrate'' (where $d_{\perp}$
is the dimension of the space transverse to the direction of
propagation), as opposed to the naive expectation of a
$d_{\perp}$-dimensional KPZ growth (i.e., two-dimensional instead
of one-dimensional in our case.) That conjecture
\cite{Saarloos_PRL01} was based on field-theoretical arguments,
but were subsequently shown by Moro \cite{Moro_01} to apply only
to systems where stochastic effects are due to external
fluctuations (such as, fluctuations in the parameters of the
model). Most recently, it was argued \cite{Moro_01,Saarloos_03}
and demonstrated \cite{Moro_01} that, in fact, fronts which in the
mean-field limit are ``pulled'', in the presence of {\em internal}
fluctuations (i.e., systems with stochastic particle dynamics),
belong to the ``usual'' $d_{\perp}$-dimensional KPZ universality
class. While corrections-to-scaling effects and system
size-limitations can often hinder a high-precision determination
of the exponents associated with interface roughening, the width
distributions provide a very strong signature and aid in
determining the universality class of the interface \cite{foltin}.
To that end, we included the two-dimensional width distribution
\cite{Marinari_PRE90} in Figs.~\ref{fig4}, supporting the
conclusion that our two-species invasion model with stochastic
particle dynamics exhibits ``standard'' (one-dimensional) KPZ
roughening, in agreement with the most recent generic arguments
\cite{Moro_01,Saarloos_03}.

As noted in the Introduction, although linear fronts are somewhat
artificial in the context of ecological invasion, they do offer
insight into more realistic scenarios. Consider, for example, that
an advantageous allele or a competitively superior species is
introduced through mutation within \cite{YKC_05,OBYKAC_05} or
through geographic dispersal to \cite{KC_JTB,OAKC_SPIE} a resident
population, respectively. Introductions occur rarely and
stochastically in both space and time. Thus, small clusters of an
advantageous allele or superior species can randomly ``nucleate''
and subsequently grow. We have shown
\cite{YKC_05,OBYKAC_05,KC_JTB,OAKC_SPIE} that the time evolution
of such systems can be well described within the framework of
homogeneous nucleation and growth. The growing clusters, on
average, have radial symmetry and reach an asymptotic velocity
$v^*$ for sufficiently long times. The corrections to the
asymptotic radial velocity of these circular fronts have two
contributions: First, the typical length of the perimeter of the
cluster scales as $L(t) \sim 2\pi R(t)\sim t$, where $R(t)$ is the
radius of the cluster. Since $z>1$, $t\ll L^{z}(t)$ for late
times, i.e., the relevant KPZ correction for radially growing
clusters is always of ${\cal O}(t^{-2/3})$ \cite{Krug_90}. Second,
for long times, when the radius of the cluster is sufficiently
large, the curvature introduces an additional ${\cal
O}(R^{-1})\sim{\cal O}(t^{-1})$ correction, subdominant to the
above KPZ correction. Thus, circular fronts are expected to reach
the same asymptotic velocity as their linear counterpart, with the
same leading-order ${\cal O}(t^{-2/3})$ corrections for late
times.

\acknowledgments
G.K. is grateful for discussions with Eli Ben-Naim and for the
hospitality of CNLS at the Los Alamos National Laboratory, where
some of this work was initiated. This research was supported in
part by the NSF through Grant Nos.\ DEB-0342689 and
DMR-0426488. Z.R. has been supported in part by the Hungarian
Academy of Sciences through Grant OTKA-T043734.

\end{document}